\documentclass[prd,aps,twocolumn,nofootinbib]{revtex4}
\usepackage{graphicx,url}
\usepackage{amsmath,amssymb}
\usepackage[utf8]{inputenc}

\def\be{\begin{equation}}
\def\ee{\end{equation}}
\def\ba{\begin{align}}
\def\ea{\end{align}}
\def\nn{\nonumber}

\def\lsim{\raise0.3ex\hbox{$\;<$\kern-0.75em\raise-1.1ex\hbox{$\sim\;$}}}
\def\gsim{\raise0.3ex\hbox{$\;>$\kern-0.75em\raise-1.1ex\hbox{$\sim\;$}}}
\def\eps{\varepsilon}
\def\theta{\vartheta}

\def\gr{gamma-ray}

\renewcommand{\vec}[1]{\boldsymbol{#1}}

\renewcommand{\vec}[1]{\boldsymbol{#1}}

\begin{document}

\title{Searching for primordial helical magnetic fields}

\author{M.~Kachelrie\ss$^{1}$}
\author{B.~C.~Martinez$^{1,2}$}

\affiliation{$^{1}$Institutt for fysikk, NTNU, Trondheim, Norway}
\affiliation{$^{2}$Astronomy Department, Wesleyan University, Middletown, CT 06459}

\begin{abstract}
The presence of non-zero helicity in intergalactic magnetic fields (IGMF) has
been suggested as a clear signature for their primordial origin. We extend
a previous analysis of diffuse Fermi-LAT gamma-ray data from 2.5 to more than
11\,years and show that a hint for helical magnetic fields in the 2.5\,year
data was a statistical fluctuation. Then we examine the detection prospects
of helical magnetic fields using individual sources as, e.g., TeV gamma-ray
blazars. We find that a detection is challenging employing realistic models
for the cascade evolution, the IGMF and the detector resolution in our
simulations.
\end{abstract}

\maketitle

%%%%%%%%%%%%%%%%%%%%%%%%%%%%%%%%%%%%%%%%%%%%%%%%%%%%%%%%%%%%%%%%%%%%%%%%%%%%%%
\section{Introduction}

Magnetic fields are known to play a prominent role for the  dynamics and
in the energy budget of astrophysical systems on galactic and smaller
scales, but their role on larger scales is still
elusive~\cite{Kulsrud:2007an,Durrer:2013pga}. 
So far, only in a few galaxy clusters observational constraints have been
obtained, either by detecting their synchrotron radiation halos or by
performing Faraday rotation measurements. Since both observational methods
need a prerequisite to measure magnetic fields (a high thermal density for
rotation measurements  and the presence of relativistic particles for radio
emission), they have been successfully 
applied only to high density regions of collapsed objects as galaxies and 
galaxy clusters. Fields significantly below the $\mu$G level are barely 
detectable with these methods. Also other constraints, for instance the 
absence of distortions  in the spectrum and the polarization properties 
of the cosmic microwave background  radiation, imply only a fairly 
large, global upper limit on the intergalactic magnetic field (IGMF) at
the level of  $10^{-9}$\,G.

An alternative approach to obtain information about the IGMF is 
to use its effect on the radiation from TeV gamma-ray sources.
The multi-TeV \gr\ flux from distant blazars is strongly attenuated 
by pair production on the infrared/optical extragalactic 
background light (EBL),  initiating electromagnetic cascades in the 
intergalactic
space~\cite{Nikishov61,Gould:1966pza,Berezinsky:1975zz,Strong:1976ea}.
The charged component of these cascades is 
deflected by the IGMF. Potentially observable effects of such 
electromagnetic cascades in the IGMF include the delayed ``echoes" of 
multi-TeV \gr\  flares or gamma-ray
bursts~\cite{Plaga:1995ins,Takahashi:2008pc,Murase:2008pe},
the appearance of extended 
emission around initially point-like \gr\ sources 
\cite{Aharonian:1993vz,Neronov:2007zz,Dolag:2009iv,Elyiv:2009bx,Neronov:2010bd}, and the suppression of GeV
halos around TeV blazars~\cite{dAvezac:2007xri}.
The last method has been used to derive lower limits on the strength and
the filling factor of the IGMF~\cite{Neronov:1900zz,Tavecchio:2010mk,Dolag:2010ni}. However, it is unclear
if plasma instabilities invalidate these claims, as argued first in
Ref.~\cite{Broderick:2011av}.

The observed magnetic fields in galaxies and galaxy clusters are assumed to
result from the amplification of much weaker seed fields. Such seeds could be
created in the early universe, e.g.\ during phase transitions or inflation,
and then amplified by plasma processes~\cite{Durrer:2013pga}. If the 
generation mechanism of such primordial fields, as e.g.\ sphalerons in the
electroweak sector of the standard model,  breaks CP then the IGMF
will have a non-zero helicity. In the case of helical fields an ``inverse
cascade'' may transfer power from smaller to larger length
scales~\cite{Brandenburg:1996fc,Olesen:1996ts}, increasing thereby its
observable effects. Moreover, helical fields decay slower than
non-helical ones. Therefore a small non-zero initial helicity fraction is
increasing with time, making the field either completely left- or
right-helical today.
%%%%
% Since helical fields decay slower than
%non-helical ones, a small non-zero initial helicity fraction is increasing
%with time, making the field either completely left- or right-helical today.
%Moreover, in the case of helical fields an ``inverse cascade'' may transfer
%power from smaller to larger length
%scales~\cite{Brandenburg:1996fc,Olesen:1996ts},
%increasing thereby its observable effects.
%%%%
A clean
signature for a primordial origin of the IGMF is therefore its non-zero
helicity. In a series of works, Vachaspati and collaborators worked out
possible observational consequences of a helical
IGMF~\cite{Tashiro:2013bxa,Tashiro:2013ita,Tashiro:2014gfa,Long:2015bda}.
Moreover, they examined gamma-ray data
from the Fermi-LAT satellite and found a 2.5\,$\sigma$ hint for the
presence of a helical IGMF~\cite{Tashiro:2013ita,Chen:2014qva}.

In this work, we re-analyze the gamma-ray data from Fermi-LAT in
Sec.~\ref{data}, extending the data set from 2.5\,years used in
Ref.~\cite{Chen:2014qva} to more than 11\,years.  We find that this data
set composed of diffuse gamma-rays is consistent with zero helicity.
This result is expected for a diffuse photon flux, because the contributions
from positive and negative charges in an electromagnetic cascade cancel in
observables sensitive to non-zero helicity. Thus a possible signal for
helical magnetic fields is suppressed in the diffuse flux, since the
number of sources per considered angular patch on the sky is typically
large. In Sec.~\ref{blazar},
we examine therefore the detection prospects of non-zero helicity in the IGMF
using individual sources as, e.g., blazars. If such a source is seen from the
side, preferentially one charge may be deflected towards the observer. Such
a charge separation may in turn render the detection of helicity possible.
Starting from a toy model similar to the one in Ref.~\cite{Long:2015bda}, we
investigate how strong the addition of realistic features like fluctuations
in the interaction length and the IGMF as well as experimental errors
deteriorates the detection prospects.
We summarize our conclusions in Sec.~\ref{concl}.

%%%%%%%%%%%%%%%%%%%%%%%%%%%%%%%%%%%%%%%%%%%%%%%%%%%%%%%%%%%%%%%%%%%%%%%%%%%%%%
\section{Diffuse gamma-rays from  10+ years of Fermi data} \label{data}

The authors of Ref.~\cite{Tashiro:2014gfa} suggested to use as observable $Q$ 
to detect a helical IGMF the triple scalar product of the arrival directions
$\vec n_i$ (normalized as $|\vec n_i|=1$) of three photons from the diffuse
gamma-ray background. Depending on their energies $\eps_i$, photons are split
into three different energy bins with lower bounds $E_3$, $E_2$, and $E_1$.
Each bin has a size $\Delta E$, given by dividing the range of photon energies
by the number of bins. Photons are binned such that
$E_i < \eps_i < E_1+\Delta E$ for $i=\{1,2,3\}$.
Photons outside these energy ranges are discarded. 
The arrival directions of the photons in the highest energy bin, i.e.\ those
with $E_3 < \eps_{3}<E_3 +\Delta E $, serve as proxy
for the direction to potential sources, since such secondary photons are
typically produced by cascade electrons with higher energies which are in
turn less deflected. If all three photons originate from the same source,
a curve connecting the highest energy photon to the lowest energy photons
in decreasing order will be bent to the right in a right-handed helical
magnetic field. Similarly, the triple scalar product
$(\vec n_1\times\vec n_2)\cdot\vec n_3$ will be positive, while it will be
negative for a left-handed helical field. The estimator $Q$ is thus defined
as~\cite{Tashiro:2014gfa}
\begin{align}
 Q = &  \frac{1}{N_{1}N_{2}N_{3}} \sum_{i,j,k}
(\vec n_i(\eps_{1})\times\vec n_j(\eps_2))\cdot\vec n_k(\eps_3)
\nn     \\ &
W_{R}(\vec n_i(\eps_1)\cdot\vec n_k(\eps_3))
W_{R}(\vec n_j(\eps_2)\cdot\vec n_k(\eps_3)),
\end{align}
where $N_i$ denotes the number of photon in the bin $i$ and $W_{R}$ a top-hat
window function. Its radius $R$ is a free
parameter  introduced to ensure that only photons within a certain angular
separation $\vec n_{i,j}\cdot\vec n_k\leq R$ of the $E_{3}$ photon
are considered. Note that the contribution of all photons which are not
actually part of the cascade should sum to zero, as they should be randomly
distributed within the patch. 

This estimator was used in Refs.~\cite{Tashiro:2013ita,Chen:2014qva} to
quantify the signatures of IGMF helicity in the diffuse gamma-ray background,
using $\approx 60$ weeks of data (August~2008 through January~2014) from
the Fermi-LAT satellite. The authors of these works reported a hint of
$2.5\sigma$ for the presence of a helical cosmological magnetic field of
left-handed helicity with
a field-strength $\sim 10^{-14}$\,G on $\sim 10$\,Mpc scales. The analysis of
Ref.~\cite{Tashiro:2013ita} included however scans over various parameters
(the limits of the energy bins $E_1$ and $E_2$, the minimally allowed Galactic
latitude $b$ for the different energy bins, and the radius $R$ of the top-hat
window function), and the true significance of this hint including penalty
factors for the ``look-elsewhere effect'' should be therefore
significantly smaller. Instead of calculating these penalty factors, 
we will use the  $\approx 60$ weeks data set as a test case to fix these cuts
which we then apply to eleven years of data (August~2008 through
September~2019).

As a first test, we check if we can reproduce the results of
Ref.~\cite{Tashiro:2013ita} by limiting the data set to events that were
detected between August 2008 and January 2014, and by applying the same
coordinate cut and scan in energy (i.e., all combinations of
$E_{1}$,$E_{2} \in\{10,20,30,40,50\}$\,GeV  with events limited to
$|b|> 60^{\circ}$). Moreover we use only events in the  Pass 8 ULTRA CLEAN class.
We are concerned with only the diffuse part of the
gamma-ray sky, so we mask out a 3$^{\circ}$ angular diameter around each
source in the first LAT high-energy catalog~\cite{Ackermann:2013fwa}. This
catalog contains 514~gamma-ray sources that were discovered in the first
three years of data-taking by Fermi-LAT. The resulting data set matches the one
of Ref.~\cite{Tashiro:2013ita} exactly, with 7,053~events in the lowest
energy bin and 200~events in the highest energy bin.

\begin{figure}[ht]
  \centering
  \includegraphics[width=0.99\columnwidth]{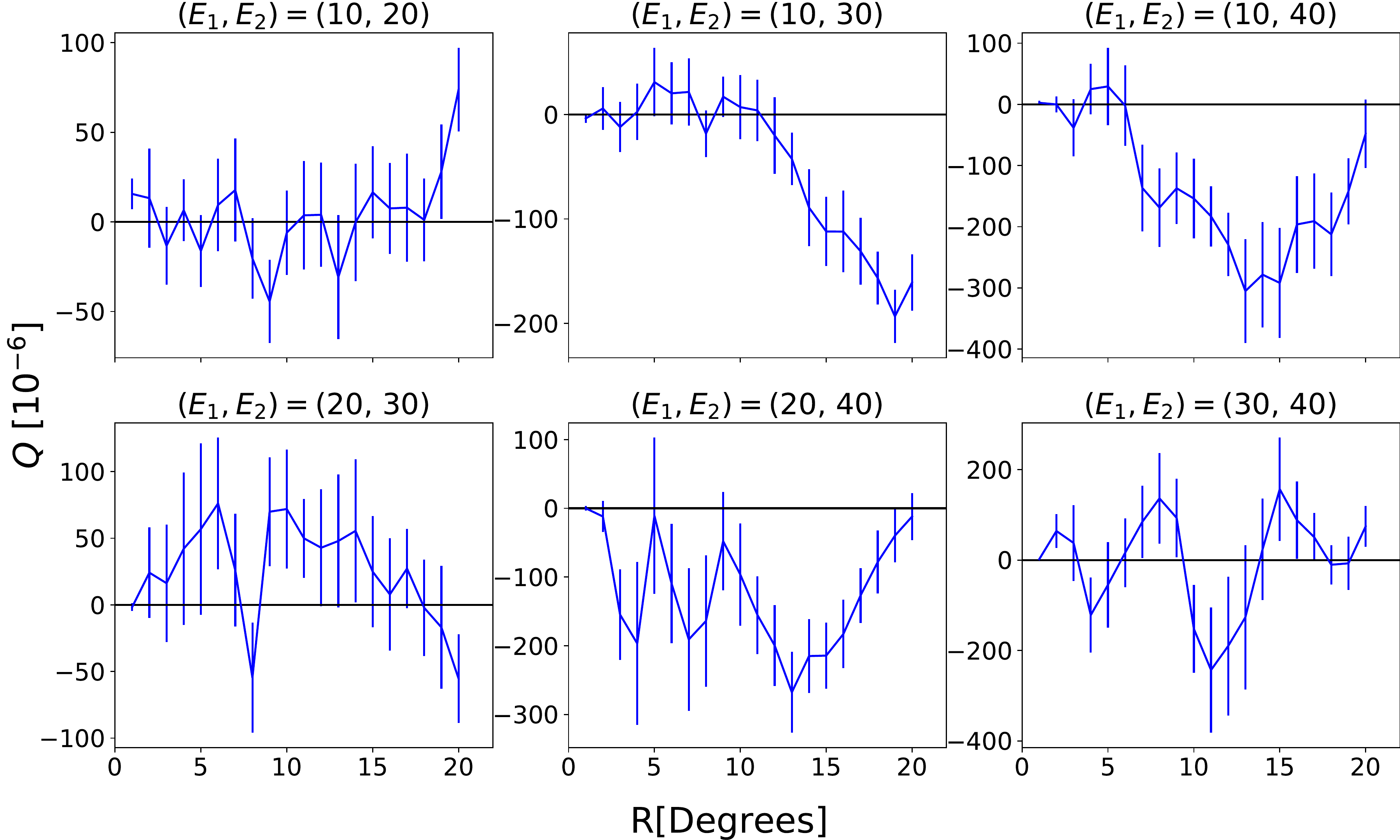}
  \caption{Estimator $Q$ as function of the opening angle $R$ of the patch for
    all combinations of $\{E_{1},E_{2}\} \in\{10,20,30,40,50\}$\,GeV. Patches are
    centered on $50 \leq \eps_{3}\leq 60$\,GeV photons with absolute Galactic latitude
    $|b|\geq 80^\circ$; data are from August 2008 through January 2014.  }
  \label{fig:Q1}
\end{figure}

In order to evaluate the $Q$ estimator, we have created our own python routines
which we have found to produce consistent results with those at \url{https://sites.physics.wustl.edu/magneticfields/wiki/index.php/Search_for_CP_violation_in_the_gamma-ray_sky}.
We have determined error bars  by calculating the standard deviation of $Q$
and dividing by $\sqrt{N}$ with $N$ is the number of $E_{3}$ events.
Figure~\ref{fig:Q1} shows the resulting $Q$ values multiplied with the
factor $10^6$.

\begin{figure}[ht]
  \centering
  \includegraphics[width=0.99\columnwidth]{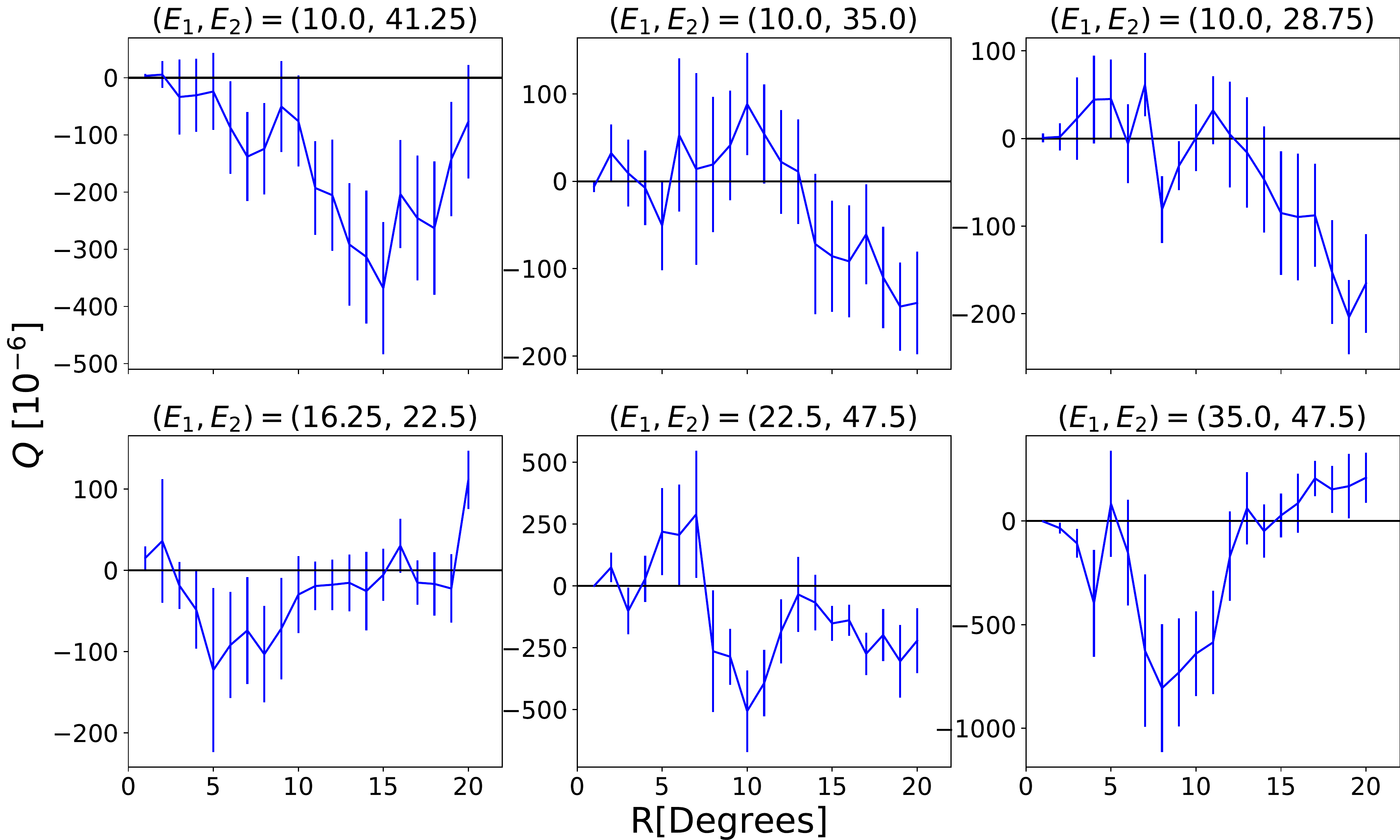}
  \caption{Estimator $Q$ as function of the opening angle $R$ of the patch for
    some combinations of $\{E_{1},E_{2}\}\in \{10,16,23,29,35,41,48,54\}$\,GeV.
    Patches are centered on $53 \leq \eps_{3}\leq 60$\,GeV photons with absolute
    Galactic latitude $|b|\geq 84^{\circ}$;
    data are from August 2008 through January 2014.}
  \label{fig:Q2}
\end{figure}

In order to find the parameters upon which the value $Q$ is maximized, we
iterate over the following free parameters: the border of the energy bins, the
minimally allowed Galactic latitude $b$ for $\eps_1$ and $\eps_2$ events,
and the
minimally allowed Galactic latitude of $\eps_3$ events. From this analysis,
we find that using for the scan eight evenly spaced values with $\eps_3$
events at Galactic latitudes $|b|\geq 84^{\circ}$ maximizes the observed
value of $Q$: Comparing Fig.~\ref{fig:Q2} to Fig.~\ref{fig:Q1}, one can see
that the $Q$ values are for some cases much larger in the
eight-binned scan.

There are currently over 11~years of gamma-ray data available from the
Fermi-LAT experiment. Applying the same cuts to this database as were initially
applied to the 2014 experiment yields a data set with 29,272 events, whereas the
old data set contained only 9,942 events. When the new data set is divided into
five evenly spaced energy bins from 10\,GeV to 60\,GeV (matching the cuts
used in Ref.~\cite{Tashiro:2013ita}), there are 20,098 events in the 10\,GeV
energy bin, 5,098 events in the 20\,GeV energy bin, 2,252 events in the
30\,GeV energy bin, 1,128 events in the 40\,GeV energy bin, and 696 events
in the 50\,GeV energy bin.

\begin{figure}[ht]
  \centering
  \includegraphics[width=0.99\columnwidth]{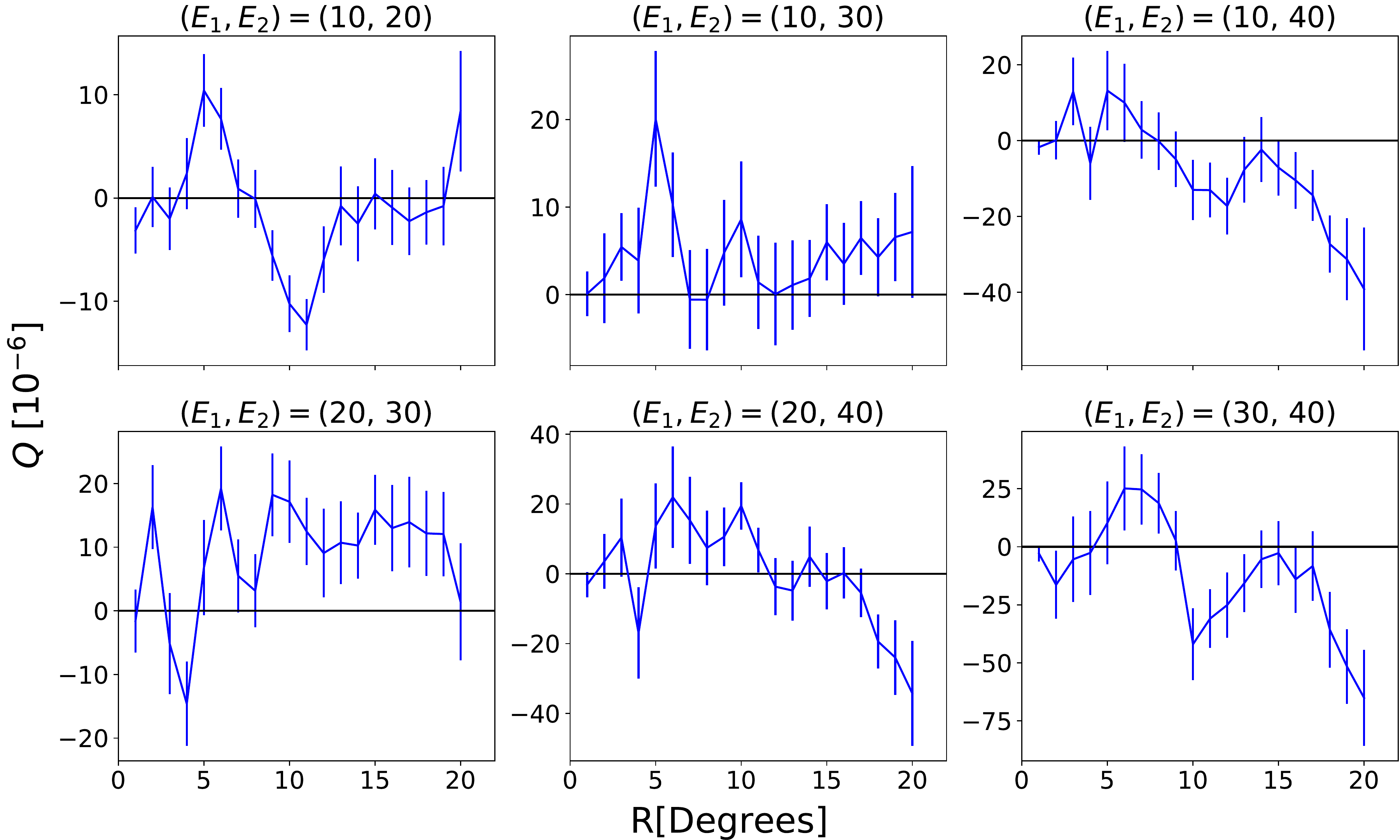}
  \caption{Estimator $Q$ as function of the opening angle $R$ of the patch for
    all combinations of $\{E1,E2\} \in\{10,20,30,40,50\}$\,GeV. Patches are
    centered on $50 \leq \eps_3\leq 60$\,GeV photons with absolute Galactic latitude
    $|b|\geq 80^{\circ}$; data are from August 2008 through September 2019.}
  \label{fig:Q3}
\end{figure}

Calculating the $Q$ estimator for this data set yields Fig.~\ref{fig:Q3}.
When comparing Fig.~\ref{fig:Q3} to Fig.~\ref{fig:Q1}, it is clear that the
$Q$ values obtained for the 2008--2014 data set are unreliable: If the
detection would have been real, one would expect to see the signal to grow
stronger in the extended data set. However, the $Q$ values are generally a
full order of magnitude smaller than those in the smaller data set.

\begin{figure}[ht]
  \centering
  \includegraphics[width=0.99\columnwidth]{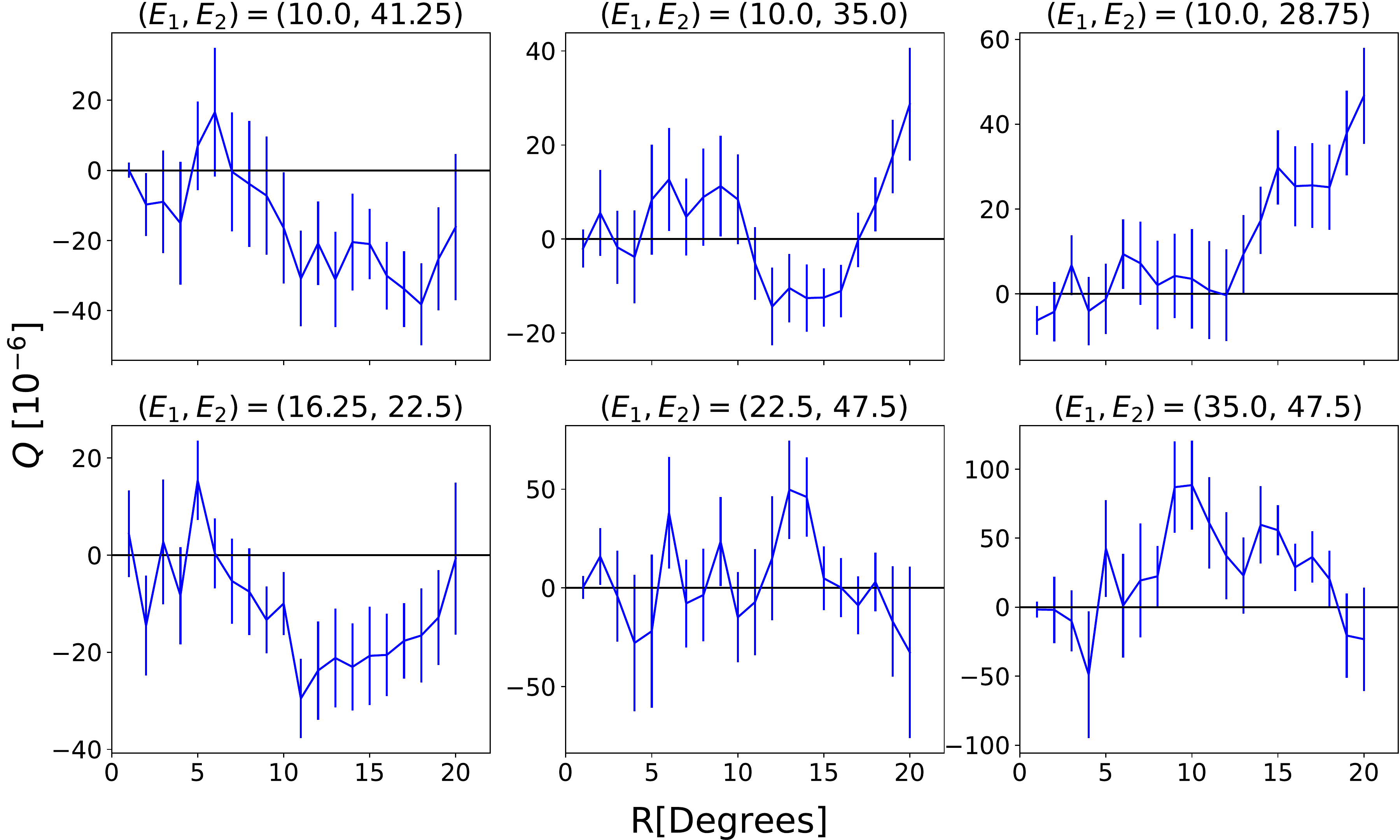}
  \caption{Estimator $Q$ as function of the opening angle $R$ of the patch for
    some combinations of $\{E1,E2\} \in\{10,16,23,29,35,41,48,54\}$\,GeV.
    Patches are centered on $53 \leq \eps_3 \leq 60$\,GeV photons with absolute
    Galactic latitude $|b|\geq 84^{\circ}$; data are from August 2008 through
    September 2019.}
  \label{fig:Q4}
\end{figure}

Applying the same analysis of the full data set using the cuts determined from
our scan in Fig.~\ref{fig:Q2} yields similar results: There is an order of
magnitude decrease in the detected signal when the full data set is analyzed.
The results of this analysis are presented in Fig.~\ref{fig:Q4}.

We conclude therefore that the hint for helical magnetic fields in the
2.5\,year  was a fluctuation. In particular, its statistical significance
was misinterpreted because the ``looking-elsewhere  effect'' was not accounted
for.

%%%%%%%%%%%%%%%%%%%%%%%%%%%%%%%%%%%%%%%%%%%%%%%%%%%%%%%%%%%%%%%%%%%%%%%%%%%%%%
\section{Gamma-rays from TeV blazars}               \label{blazar}

In this section, we investigate if extended halos of TeV blazars can
be used for the detection of helical magnetic fields. This question
was addressed in a series of works using mainly analytical toy
models~\cite{Tashiro:2013bxa,Tashiro:2013ita,Tashiro:2014gfa,Long:2015bda,AlvesBatista:2016urk}:
Neglecting in particular a continuous injection spectrum, fluctuations in the
distribution of interaction lengths and the energy distribution of secondaries, it was shown
that helical magnetic fields can be detected, if the distance to the
source and the correlation length of the turbulent field satisfies
certain criteria. Here we use instead of such analytical toy models the
version~3.03 of the Monte Carlo program
{\tt ELMAG}~\cite{Kachelriess:2011bi,Blytt:2019xad} to simulate the
three-dimensional evolution of electromagnetic cascades. Replacing in this
program the probability distribution functions (pdf) usually employed by
their expectation
values, we can emulate the analytical toy models used previously. Switching on
these sources of fluctuations, we can identify those which affect mostly
the detection of helical magnetic fields and understand how large the
detection prospects are under realistic assumptions.

If not otherwise specified, we use a source located at the redshift $z=0.25$,
corresponding to the comoving radial distance $\simeq 1$\,Gpc. We assume as
opening angle of the blazar jet  $\Theta_{\rm jet}=1^\circ$  and inject 10,000
photons into a magnetic field of strength $B_{\rm rms} = 10^{-16}$\,G.  To
calculate $Q$, we use photons of energy $10< \eps_{1}/{\rm GeV} < 26$ and
$74< \eps_{2}/{\rm GeV}<90$ in bins of size $\Delta E = 16$\,GeV.
As EBL model, we choose the one of Gilmore {\it et al.}~\cite{Gilmore:2011ks}.

In order to quantify the detectability of helicity at each stage, we calculate
the distribution of $Q$ values for typically $N=500$ Monte Carlo sets. Since
we know the position of the source, we adapt the algorithm used in section~II
slightly, substituting the highest-energy photons $\eps_{3}$, which would only approximate the position of the source,  with the actual position of
the source. Thus we use only photons produced as secondaries in the cascades
for $E_{1}$ and $E_{2}$. Since we do not include background photons in our
simulations, it is favorable to use no angular cuts and to include thereby
as many photons as possible in the calculation of $Q$.  

We find that the probability distribution of $Q$ values is rather
well-described by a Gaussian (or normal) distribution, with the deviations
increasing as we add more and more sources of physical fluctuations. For each
parameter set, we fit two Gaussians $N(\mu_i,\sigma_i^2)$ to the two
distributions of $Q$ values for the right- and left-helical magnetic
fields and calculate then their overlap $O$,
\be \label{O}
O= 1-\frac12\mathrm{erf}\left(\frac{c-\mu_1}{\sqrt2\sigma_1}\right)
 +\frac12\mathrm{erf}\left(\frac{c-\mu_2}{\sqrt2\sigma_2}\right) ,
\ee
where $\mu_1<\mu_2$ and $c$ is the intersection of the two distributions.
%
%\be
% c = \frac{\mu_2\sigma_1^2-\sigma_2\left\{\mu_1\sigma_2+\sigma_1[(\mu_1-\mu_2)^2+2(\sigma_1^2-\sigma_2^2)\ln(\frac{\sigma_1}{\sigma_2})]^{1/2}\right\}}{\sigma_1^2-\sigma_2^2}
%\ee
%
When the percentage of overlap between the two distributions is high, positive
and negative helical magnetic fields are difficult to distinguish using the
$Q$ distribution. Similarly, one can use the overlap to quantify how
likely the ``signal hypothesis'' $|h|=1$ can be distinguished from
the ``background hypothesis'' $|h|=0$.

%%%%%%%%%%%%%%%%%%%%%%%%%%%%%%%%%%%%%%%%%%%%%%%%%%%%%%%%%%%%%%%%%%%%%%%%%%%%%%
\subsection{Toy model}

As a first step, we test the ability to detect with {\tt ELMAG} magnetic
field helicity in the highly idealized scenario adapted from
Ref.~\cite{Long:2015bda}.
In this toy model, photons are injected with a fixed energy $E=10^{13}$\,eV.
The magnetic field consists of a single mode with helicity
$h=\{-1,0,1\}$,
\be \label{one}
\vec B(\vec r)= \left( \begin{array}{c}
                         B\sin(kz+\beta) \\
                         hB\cos(kz+\beta) \\
                         0
                       \end{array}  \right)
\ee
and its wave-vector $\vec k$ is pointing along the line-of-sight towards the
blazar. Moreover, we replace the pdf for the interaction lengths by their
expectation values. Thus the only source of fluctuations in this toy model
are the energy fractions transferred to the secondary particles.
Finally, we switch off the creation of one charge
in the process $\gamma\gamma\to e^+e^-$ to create by hand
a charge asymmetry. We set the observation angle $\Theta_{\rm obs}$ between
the line-of-sight and the jet axis to zero. 

\begin{figure*}[ht]
  \centering
  \includegraphics[width=2\columnwidth]{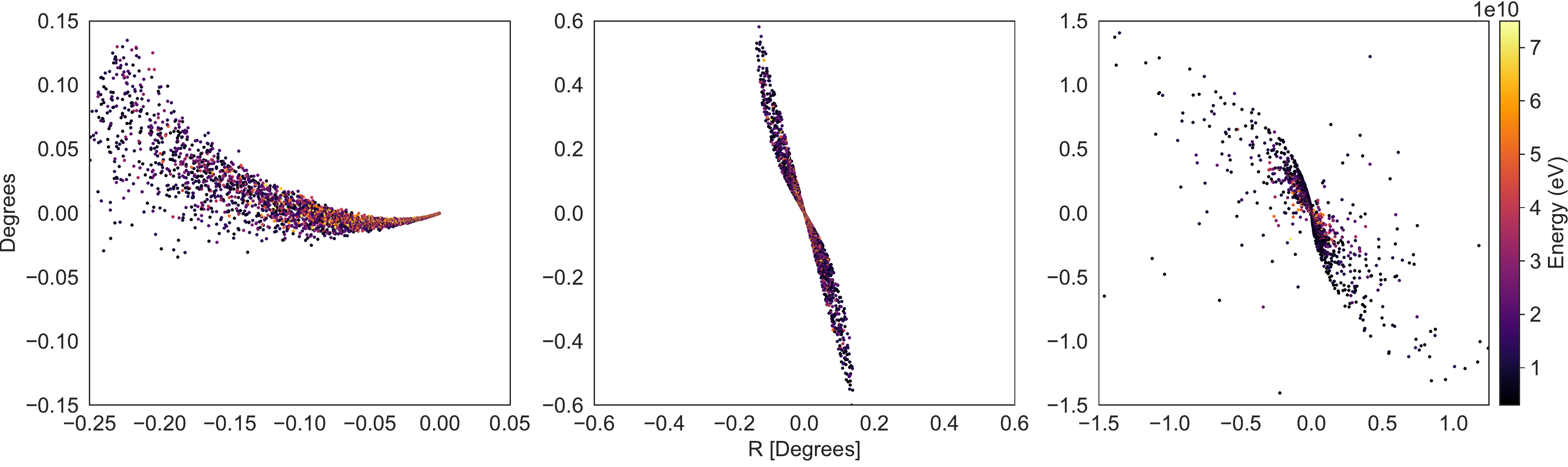}
  \caption{Sky map of the arrival directions of photons: left is the toy model with  $L_{\max}=0.1$\,Mpc, the middle panel includes both charges and continuous injection spectrum with $L_{\max}=600$\,Mpc, and the right is for the realistic case including PDF for interaction lengths,  3D magnetic turbulence and $L_{\max}=600$\,Mpc.}
  \label{fig:skymap}
\end{figure*}

\begin{figure}[ht]
  \centering
  \includegraphics[width=0.99\columnwidth]{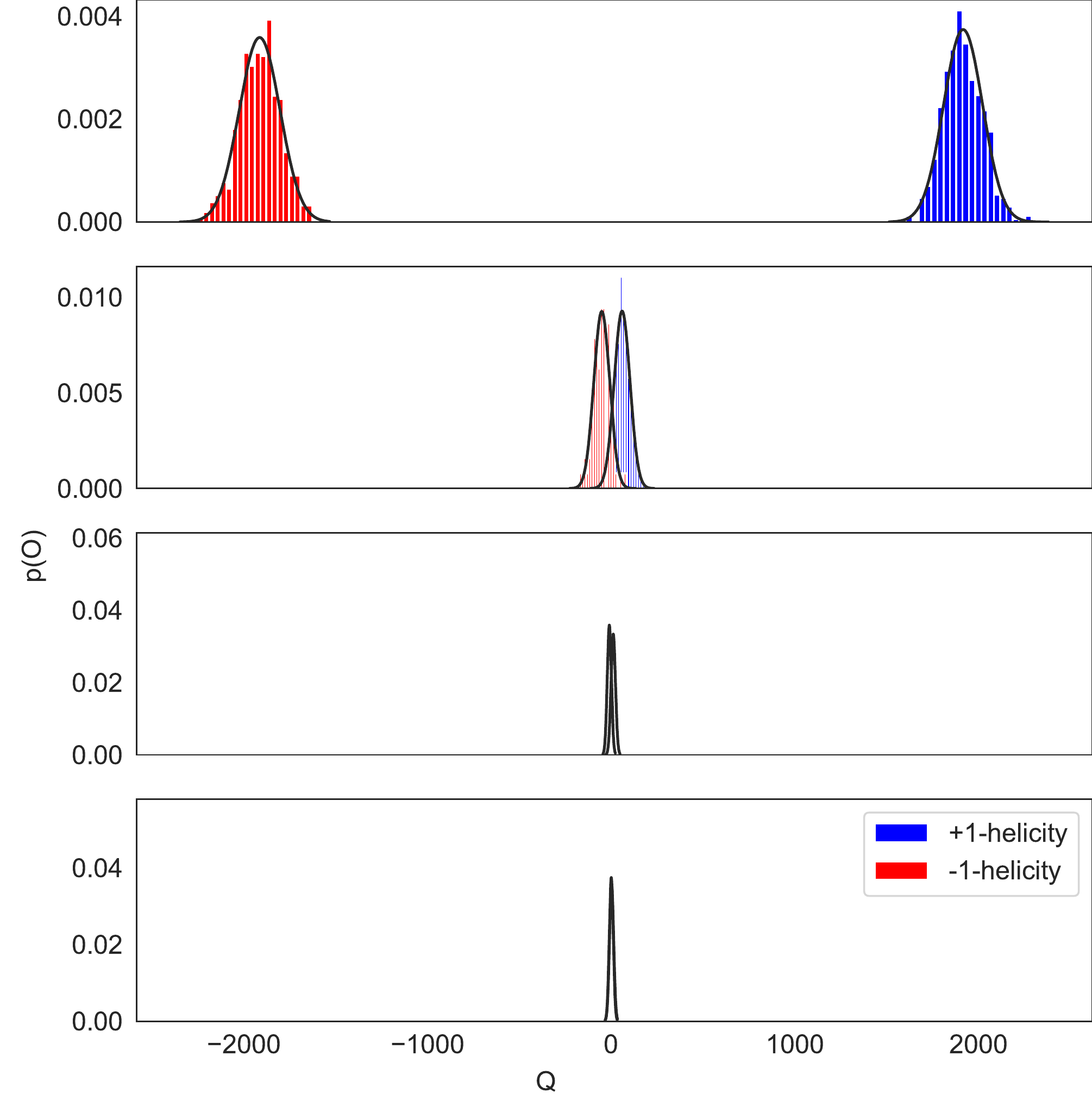}
  \caption{MC results and fitted normal distributions of $Q$ values for the
    toy model, orange bins for $h=-1$ and blue bins for $h=+1$;
    the wave-lengths $L$ of the magnetic field mode are, from top to bottom,
    0.1\,Mpc, 1\,Mpc, 10\,Mpc, and 100\,Mpc.}
  \label{fig:Corr_rel_1}
\end{figure}

In the left panel of Fig.~\ref{fig:skymap}, we show the sky map of the arrival
directions of photons for $L=2\pi/k= 0.1$\,Mpc
of the magnetic field: A spiral pattern is clearly visible, with the random
phase $\beta$ leading only to a rotation of the pattern. 
Next we estimate the possibility of differentiating between a right and a
left-helical IGMF, using as measure the overlap between the probability
distributions for the $Q$ values in the two cases.
In Fig.~\ref{fig:Corr_rel_1}, we show the Monte Carlo (MC) results
together with the fitted normal distributions for various
wave-lengths of the single magnetic field mode, from $L=0.1$\,Mpc to
100\,Mpc, using 500 simulations. The ability to detect a helical
IGMF is highly dependent on the wave-length, since the field resembles more
and more a uniform one as $L$ increases: As a result, the distance
between the peaks of the normal distributions decreases and their overlap
increases. Using (3), we calculate an overlap of $1.27 \times 10^{-66}$ for a wave-length of $L= 0.1$\,Mpc, meaning that helicity is perfectly
detectable. For $L = 1$\,Mpc, the overlap of the two peaks is only 19.9\%,
increasing to 35.2\% for $L= 10$\,Mpc. Finally, the signal for helicity
practically disappears for $L=100$ Mpc, where the overlap increases to 93.5\%.

%%%%%%%%%%%%%%%%%%%%%%%%%%%%%%%%%%%%%%%%%%%%%%%%%%%%%%%%%%%%%%%%%%%%%%%%%%%%%%
\subsection{Towards the realistic case}

%%%%%%%%%%%%%%%%%%%%%%%%%%%%%%%%%%%%%%%%%%%%%%%%%%%%%%%%%%%%%%%%%%%%%%%%%%%%%%
\subsubsection{Adding continuous injection spectrum}

We now modify our highly idealized toy model to include a continuous injection
spectrum, but keep otherwise all parameters unchanged. Inspired by
high frequency peaked blazars as 1ES~0229+200, we use as injection
spectrum the power law $dN/dE_{\rm inj}\propto E_{\rm inj}^{-2/3}$ between
$10^{11} \leq E_{\rm inj} \leq 10^{13}$\,eV. Including a spectrum of injection
energies weakens the correlation between deflection angle and energy, on
which the estimator $Q$ is based on. Therefore one may expect that the
detection of a helical IGMF becomes more difficult.

\begin{figure}[ht]
  \centering
  \includegraphics[width=0.99\columnwidth]{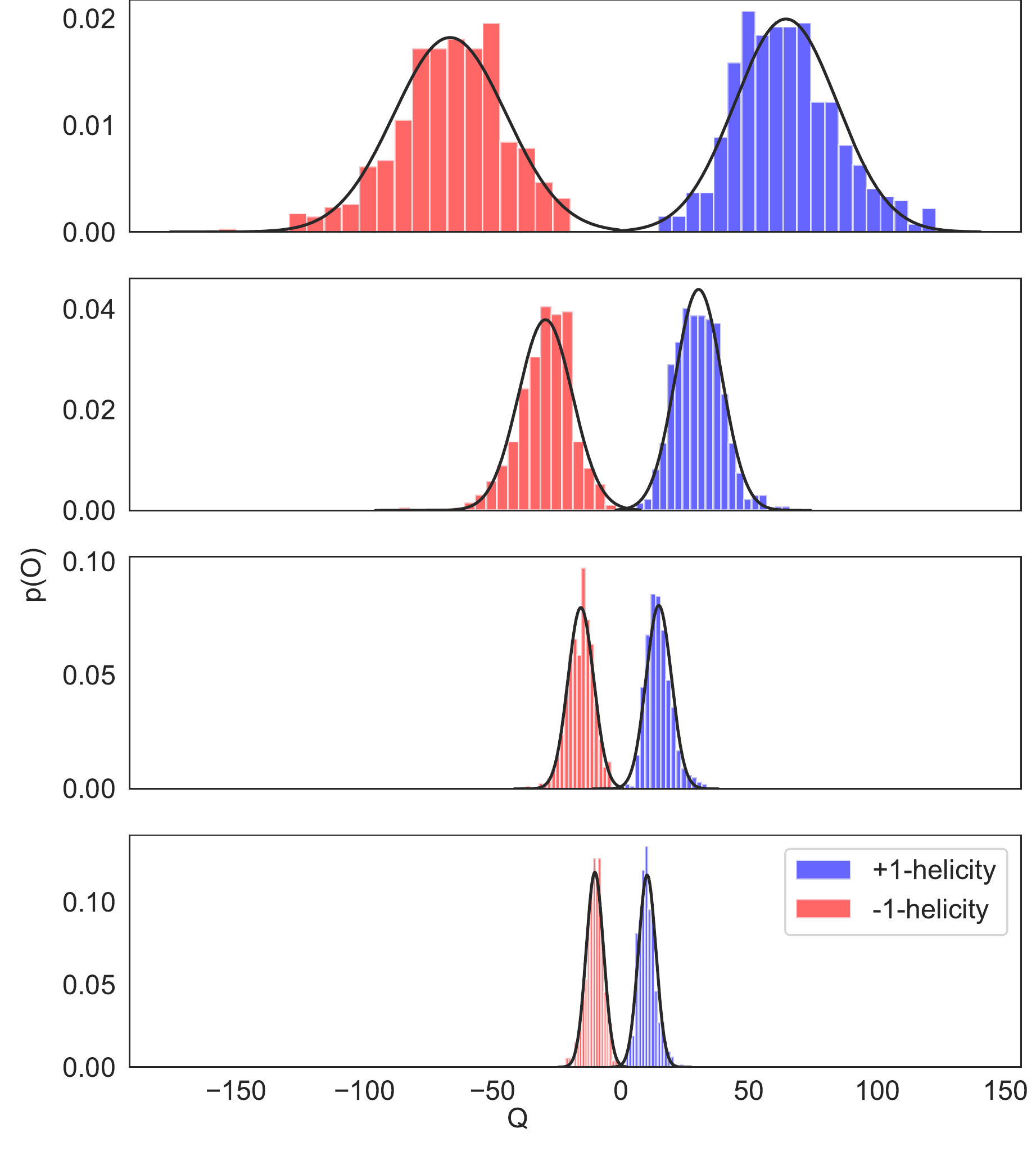}
  \caption{MC results and fitted normal distributions of $Q$ values for toy model at various correlation lengths, orange bins for $h=-1$ and blue bins for $h=+1$. Photons are injected on a continuous spectrum of energies by a source located at 1\,Gpc. The correlation lengths for each panel, from top to bottom are: 100\,Mpc, 250\,Mpc, 500\,Mpc, and 750\,Mpc. The overlap in each case is approximately 0.}
  \label{fig:Corr_rel_2}
\end{figure}

In Fig.~\ref{fig:Corr_rel_2}, we show the MC results and the fitted normal
distributions of $Q$ values obtained choosing $L=100$, 250, 500, and 750\,Mpc
for the wave-length of the magnetic field mode~(\ref{one}). Using $N=500$
Monte Carlo simulations, there is no overlap in any case,
so that
helicity is detectable with a confidence level (C.L.) of at least $99.8\%$.
Assuming Gaussian statistics and applying Eq.~(\ref{O}), the C.L. are
99.8\% for $L=100$\,Mpc, 99.8\% for $L=250$\,Mpc, 99.8\% for $L=500$\,Mpc, and
99.7\% for $L=750$\,Mpc, corresponding to a $3\sigma$ detection using a single
source for each case. 

These results disagree with the expectation that helicity becomes more
difficult to detect using a continuous energy spectrum for the injected
photons. The apparent contradiction is resolved noting that the relevant
scale to which the wave-length $L$ of the field mode should be compared
is the Larmor radius
\be
 r_{\rm L} = \frac{\gamma m v_\perp}{|q|B}
\ee
of the produced cascade electrons: It is the dimensionless ratio
$L/r_{\rm L}$ which controls if the helical magnetic field appears
effectively as a uniform field for the propagating electron.
Since the typical energy of these
electrons is much lower using a photon injection spectrum extending down
to $10^{11}$\,eV, the relevant dimensionless ratio $L/r_{\rm L}$ is increased.
Thus larger values of $L$ lead in the case of the continuous spectrum to
comparable results to the fixed energy case.

%%%%%%%%%%%%%%%%%%%%%%%%%%%%%%%%%%%%%%%%%%%%%%%%%%%%%%%%%%%%%%%%%%%%%%%%%%%%%%
\subsubsection{Adding both charges}

\begin{figure}[ht]
  \centering
  \includegraphics[width=0.99\columnwidth]{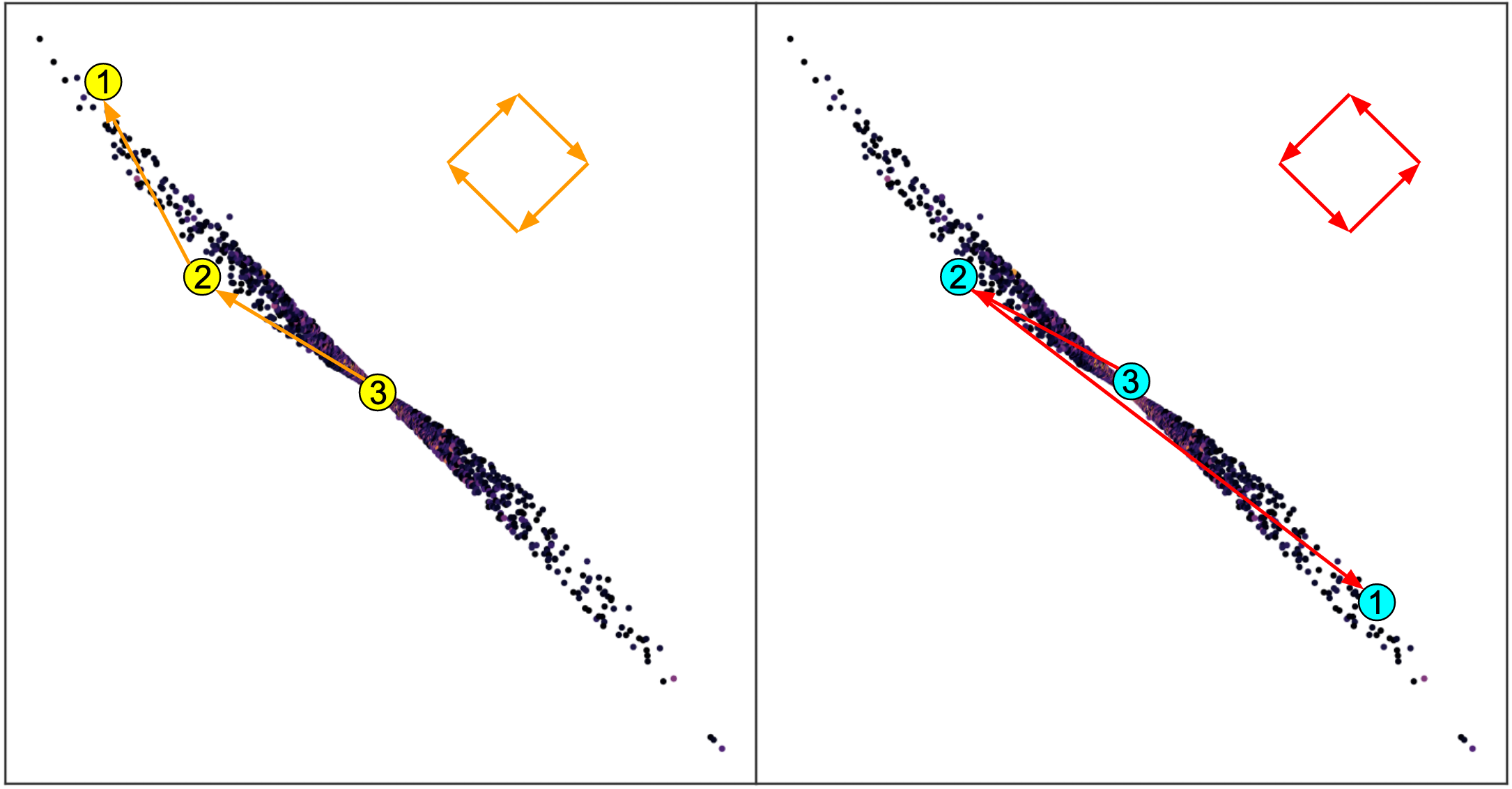}
  \caption{Illustration how the combination of $E_1$ and $E_2$
    photons from an electron and a positron cascade leads to a contribution
    to $Q$ with the wrong sign.}
  \label{fig:ilu}
\end{figure}

\begin{figure}[ht]
  \centering
  \includegraphics[width=0.99\columnwidth]{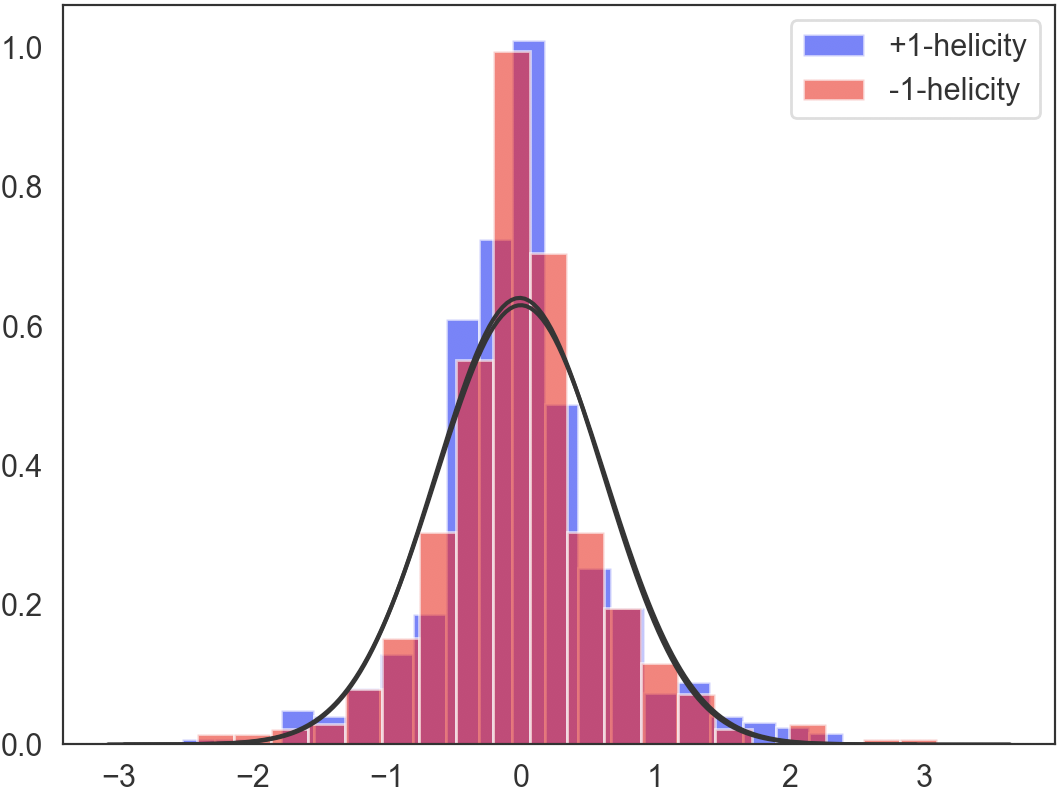}
  \caption{MC results and fitted normal distributions of $Q$ values for the
    toy model with continuous injection spectrum observed at
    $\Theta_{\rm obs}=0^\circ$, orange bins for $h=-1$, blue bins for
    $h=+1$, and red for bins with overlap;; The percentage of overlap for the two distributions is 98\%. }
  \label{fig:PL_avg_0_defl}
\end{figure}

We continue to use the continuous injection spectrum and the single magnetic
field mode~(\ref{one}), but include now both charges created in the process
$\gamma\gamma\to e^+e^-$, thereby removing the artificial charge asymmetry. 
As a result, helicity cannot be detected when the source is observed directly
face on: As shown in the middle panel of Fig.~\ref{fig:skymap}, the arrival
directions of photons originating from both electron and positron cascades
are distributed along anticlockwise spirals in a magnetic field with negative
helicity. Thus using photons from either electron or positron cascades leads
to a negative contribution to $Q$. By contrast, combining $E_1$ and $E_1$
photons from an electron and a positron cascade leads to a clockwise spiral
as shown in Fig.~\ref{fig:ilu} and, thus, to a positive contribution to $Q$. 
As a result, the various contributions to $Q$ cancel, leading to a probability
distribution consistent with zero helicity\footnote{We have tested that this
  result is not restricted to the case that $\vec k$ is parallel to the
  line-of-sight, but holds also for realistic magnetic fields with a
  three-dimensional spectrum of $\vec k$ modes.}, cf.\ with
Fig.~\ref{fig:PL_avg_0_defl}.
The source must therefore be observed off angle, so that either electrons
or positrons are deflected preferentially towards the observer.

The bending angle of a cascade under the influence of an IGMF is approximately
given by~\cite{Tashiro:2013bxa}
\begin{align} \label{theta}
\Theta(E_{\gamma}) & \approx 0.004^{\circ} \biggl( \frac{B}{10^{-16}\mathrm{G}}\biggr)\biggl(\frac{1\mathrm{Gpc}}{D_{s}}\biggr)\biggl(\frac{E_{\gamma}}{100\mathrm{GeV}}\biggr)^{-3/2},
\end{align}
so the size of the source on the sky increases linearly with the strength of
the magnetic field.  A stronger magnetic field, then, means that there is a
larger range of $\Theta_{\rm obs}$ for which helicity can be detected. 
Therefore, we slightly modify our standard parameters to better highlight
the conditions
that make detection favorable: We now assume an opening angle for the blazar
jet $\Theta_{\rm jet}=2^\circ$ and inject photons into a magnetic field of
strength $B = 10^{-15}$\,G.

\begin{figure}[ht]
  \centering
  \includegraphics[width=0.99\linewidth]{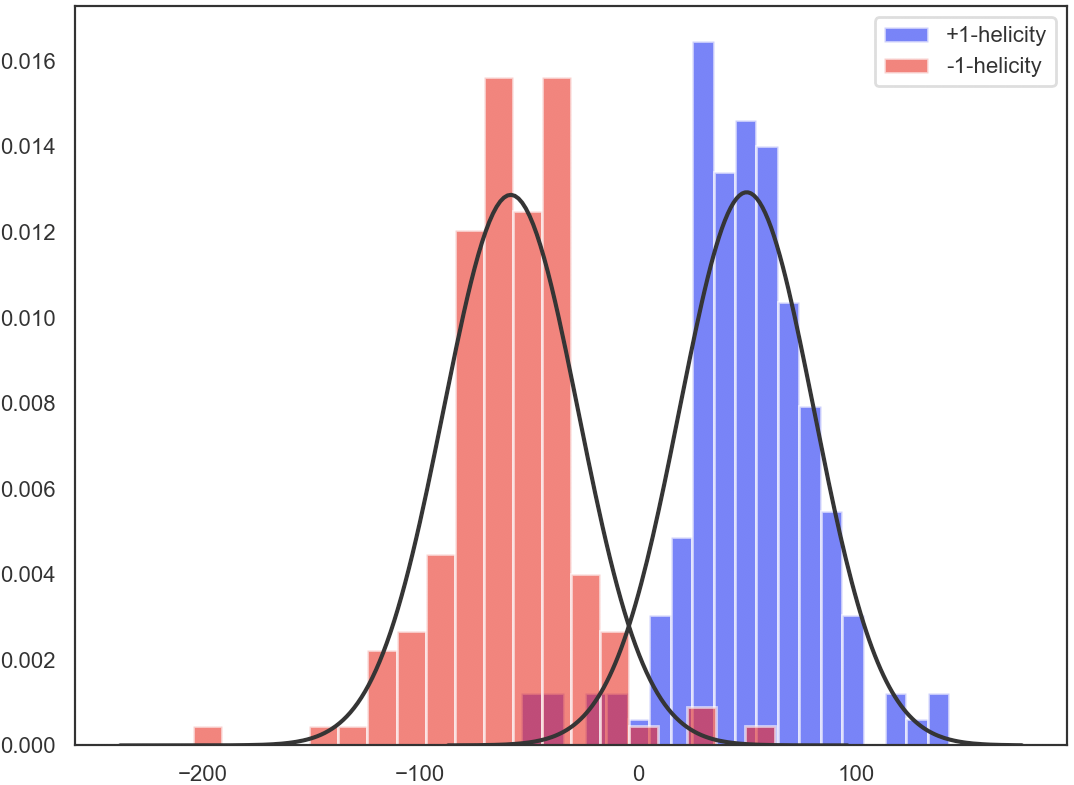}
  \caption{MC results and fitted normal distributions of $Q$ values for a
    source located at 1\,Gpc and observed at $\Theta_{\rm obs}=2.5^{\circ}$,
    orange bins for $h=-1$, blue bins for $h=+1$, and red for bins with
    overlap; both charged are included.}
  \label{fig:PL_avg_25_defl}
\end{figure}

Increasing the observation angle $\Theta_{\rm obs}$ leads to a stronger signal
until a maximum observation angle is reached, beyond which the number of
photons detected drops fast. As an example, Fig.~\ref{fig:PL_avg_25_defl}
shows the $Q$  distributions for a source located at 1\,Gpc and observed
at $\Theta_{\rm obs}=2.5^{\circ}$, i.e.\ slightly larger than the jet opening
angle  $\Theta_{\rm jet}=2^\circ$. In this case, one charge is preferentially
deflected towards the observer, and as a result, the contributions of
cascade electrons and positrons to $Q$ do not cancel out. In this specific
case, the overlap between the two distributions is only 8\%.
In Fig.~\ref{fig:PL_obs_ang_OLAP_1}, we show  the overlap between the
$Q$ distributions for a left- and right-helical fields as function of
$\Theta_{\rm obs}$ as blue curve. For small $\Theta_{\rm obs}$, the presence
of photons from opposite charge cascades weakens the helicity signal,
while for $\Theta_{\rm obs}\gsim \Theta_{\rm jet}$, the overlap
goes to zero.
We conclude that, provided that the source is observed at a sufficiently
large $\Theta_{\rm obs}$, charges are separated well enough to not destroy
the possibility to detect helical magnetic fields.

Note however that, as $\Theta_{\rm obs}$ increases, it becomes less likely
that photons reach the observer. Thus a larger number of sources is
needed to produce statistically significant data sets.

%%%%%%%%%%%%%%%%%%%%%%%%%%%%%%%%%%%%%%%%%%%%%%%%%%%%%%%%%%%%%%%%%%%%%%%%%%%%%%
\subsubsection{Adding fluctuations in the interaction length}

Keeping all other properties the same, we now replace the mean-values for
the interaction lengths of the reaction $\gamma+\gamma\to e^++e^-$ and
$e^\pm+ \gamma\to e^\pm + \gamma$ by their pdf's. Fluctuations in these
interaction lengths weaken the relation between the  energy
and the deflection angle of the observed photons. Therefore we expect
that the detection prospects of helical fields will be reduced.

\begin{figure}[ht]
  \centering
  \includegraphics[width=0.99\linewidth]{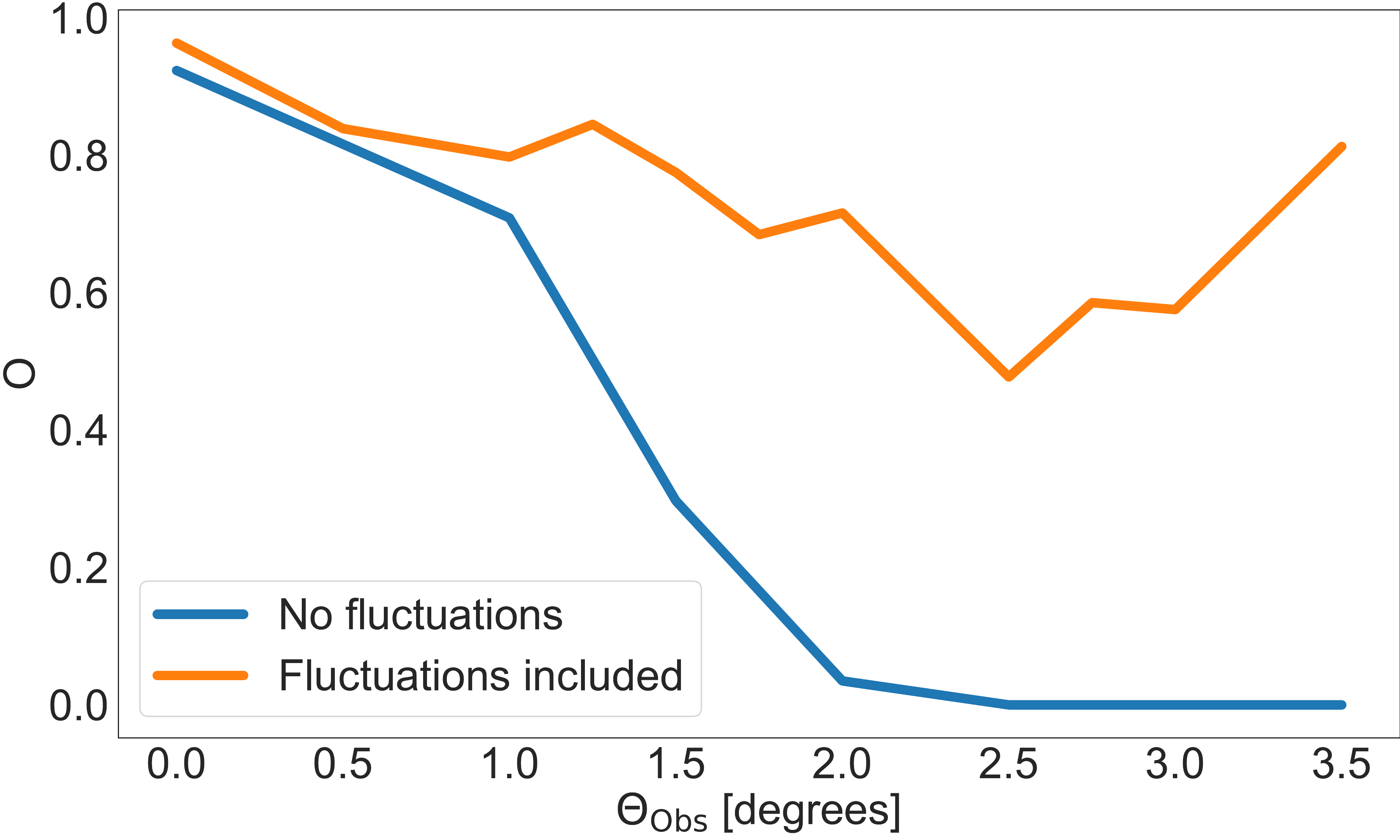}
  \caption{Overlap $O$  as a function of the observation angle
    $\Theta_{\rm obs}$; the blue curve uses the mean value, the orange curve
    the pdf's for the interaction lengths.}
  \label{fig:PL_obs_ang_OLAP_1}
\end{figure}

\begin{figure}[ht]
  \centering
   \includegraphics[width=0.99\linewidth]{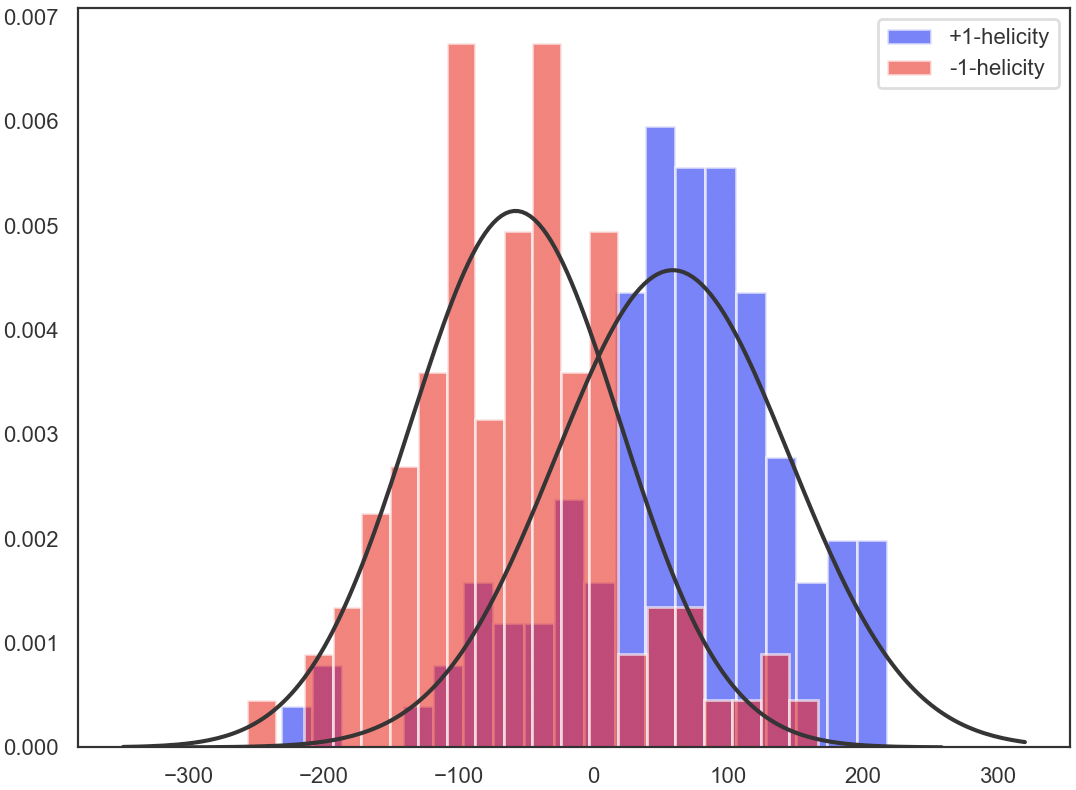}
  \caption{MC results and fitted normal distributions of $Q$ values for a
    source located at 1 Gpc and observed at $\Theta_{\rm obs}= 2.5^{\circ}$
    with an opening angle of $\Theta_{\rm jet}= 2.0^{\circ}$,
    orange bins for $h=-1$ and blue bins for $h=+1$;
    fluctuations in the interaction length are included.}
  \label{fig:Spec_Eng_def_2_5}
\end{figure}

Repeating the previous analysis, we find now that the overlap of the $Q$
distributions shown in Fig.~\ref{fig:Spec_Eng_def_2_5} increases from 8\%
to 47\%.  The orange curve in Fig.~\ref{fig:PL_obs_ang_OLAP_1} shows the
overlap as function of $\Theta_{\rm obs}$ including fluctuations: The signal
is considerably weaker, but the same trends are followed until
$\Theta_{\rm obs} \approx 2.5^{\circ}$. For $\Theta_{\rm obs} > 2.5^{\circ}$,
however, the overlap begins to increase again.

%%%%%%%%%%%%%%%%%%%%%%%%%%%%%%%%%%%%%%%%%%%%%%%%%%%%%%%%%%%%%%%%%%%%%%%%%%%%%%
\subsubsection{Adding a spectrum of $B$ modes}

\begin{figure}[ht]
  \centering
  \includegraphics[width=0.99\linewidth]{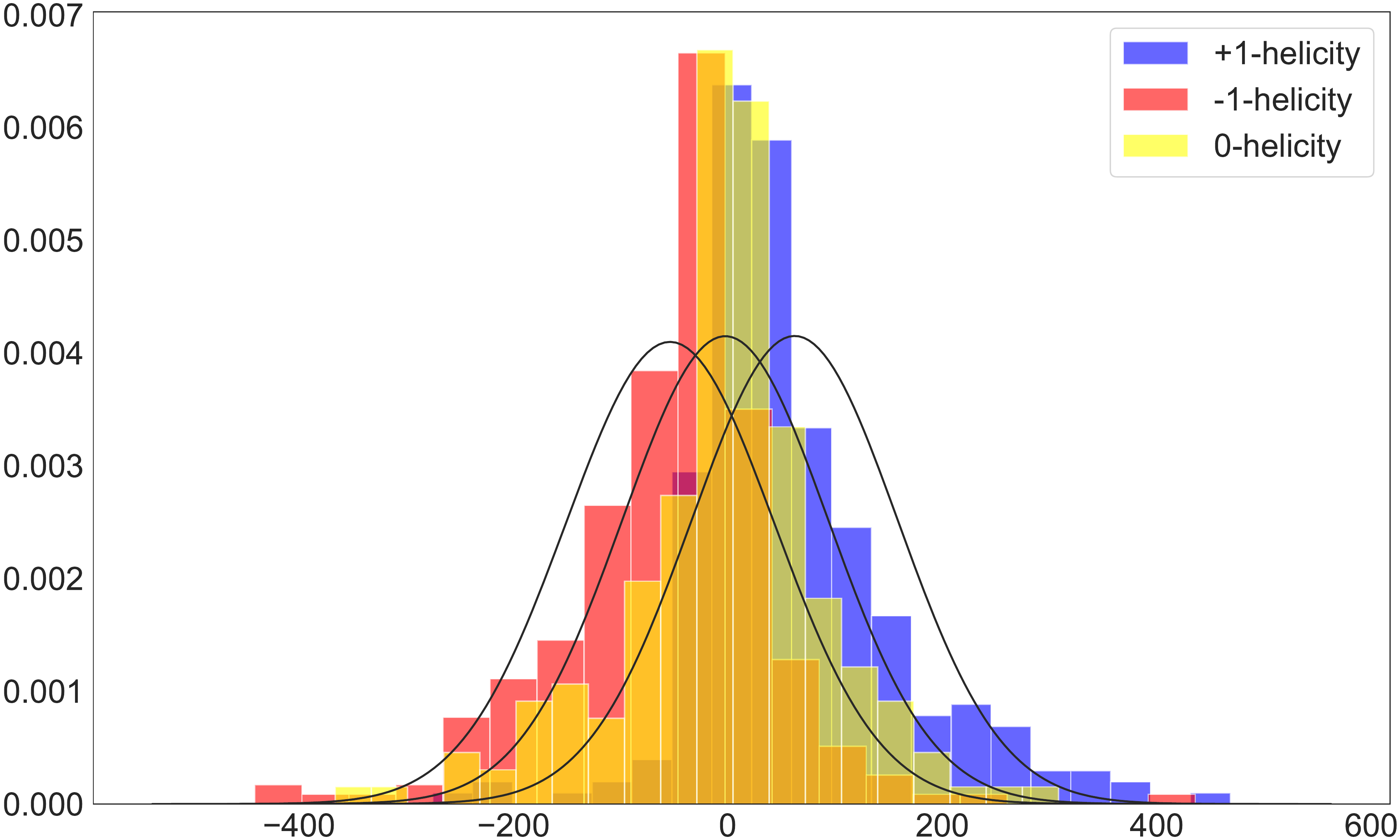}
  \caption{MC results and fitted normal distributions of $Q$ values for a
    source located at 1 Gpc, observed at $\Theta_{\rm obs}$ = 2.5$^{\circ}$ with an opening angle of $2.0^{\circ}$, orange bins for $h=-1$, yellow bins for $h=0$, and blue bins for $h=+1$. Photons are injected into a turbulent magnetic field with 100 3D modes.}
  \label{fig:3Dturb}
\end{figure}

We now include additional modes of the magnetic fields, but keep all other
properties of the simulation the same. The directions of the wave-vectors
$\vec k_i$ of each mode $i$ are distributed uniformly on the sphere $S^2$,
while their norm is distributed according to a Kolmogorov power-spectrum. In
order to keep the computing times manageable, we use maximally 100~modes.
In the right panel of Fig.~\ref{fig:skymap}, we show the sky map of the arrival
directions of photons for a maximal length of the magnetic field modes
$L_{\max}=600$\,Mpc: The spiral pattern has become rather fuzzy, indicating
that helicity becomes more difficult to detect.
The resulting $Q$ distributions are shown for the same parameters as in the
previous plot in  Fig.~\ref{fig:3Dturb}. The overlap between the $h = +1$ and
$h = -1$ pdf's increases mildly to $\approx 53\%$ and thus including a
distribution of field modes does not deteriorates significantly the detection
prospects of helical magnetic fields. Also included in Fig.~\ref{fig:3Dturb}
is the distribution for the case that $h=0$, which is as expected centered
approximately at $Q=0$. The overlap $O$ between the $h=|1|$ and $h=0$
distribution is $75\%$.

Let us comment now on the statistical significance, if $N$ such sources
are observed.  Assuming a Gaussian distribution of the $Q$ values with
an overlap of 75\%, we can use a Kolmogorov-Smirnov test to estimate
the chance probability that for $N$ observations the two distributions
are confused: While a $3\sigma$ evidence for helical fields would require
$N\simeq 50$ sources, a $5\sigma$ detection would need the observation of
$N\simeq 120$ sources with such a signal.

%%%%%%%%%%%%%%%%%%%%%%%%%%%%%%%%%%%%%%%%%%%%%%%%%%%%%%%%%%%%%%%%%%%%%%%%%%%%%% 
\subsubsection{Experimental constraints}

We have not taken into account yet constraints like the energy threshold,
the finite angular resolution and the limited observation time of specific
experiments. While a detailed discussion of these issues is outside the scope
of this work, we comment briefly on the case of the Cherenkov Telescope Array
(CTA).

The angular resolution of the Southern Array of CTA is estimated to vary
between $\ell_{68}=0.05^\circ$ at 1\,TeV and $0.15^\circ$ at 100\,GeV, where
$\ell_{68}$ denotes the angular opening angle of cone containing 68\% of
all reconstructed photons~\cite{Maier:2019afm}.
This value is in a considerable part of the relevant $\{B,D_s\}$ parameter
space comparable to or larger than the extension of the
gamma-ray halo given by Eq.~(\ref{theta}). Only for magnetic field
strengths $B\gsim 10^{-14}$\,G, we expect that the spiral patterns
contained in the halo of TeV blazars are sufficiently large such that
they are not washed out by the finite angular resolution. 
Another effect deteriorating the detection prospects for helical fields
is the reduced sensitivity of CTA below 100\,GeV, which requires an
adjusting of the energy cuts $E_1$ and $E_2$.

\begin{figure*}[ht]
  \centering
  \includegraphics[width=2\columnwidth]{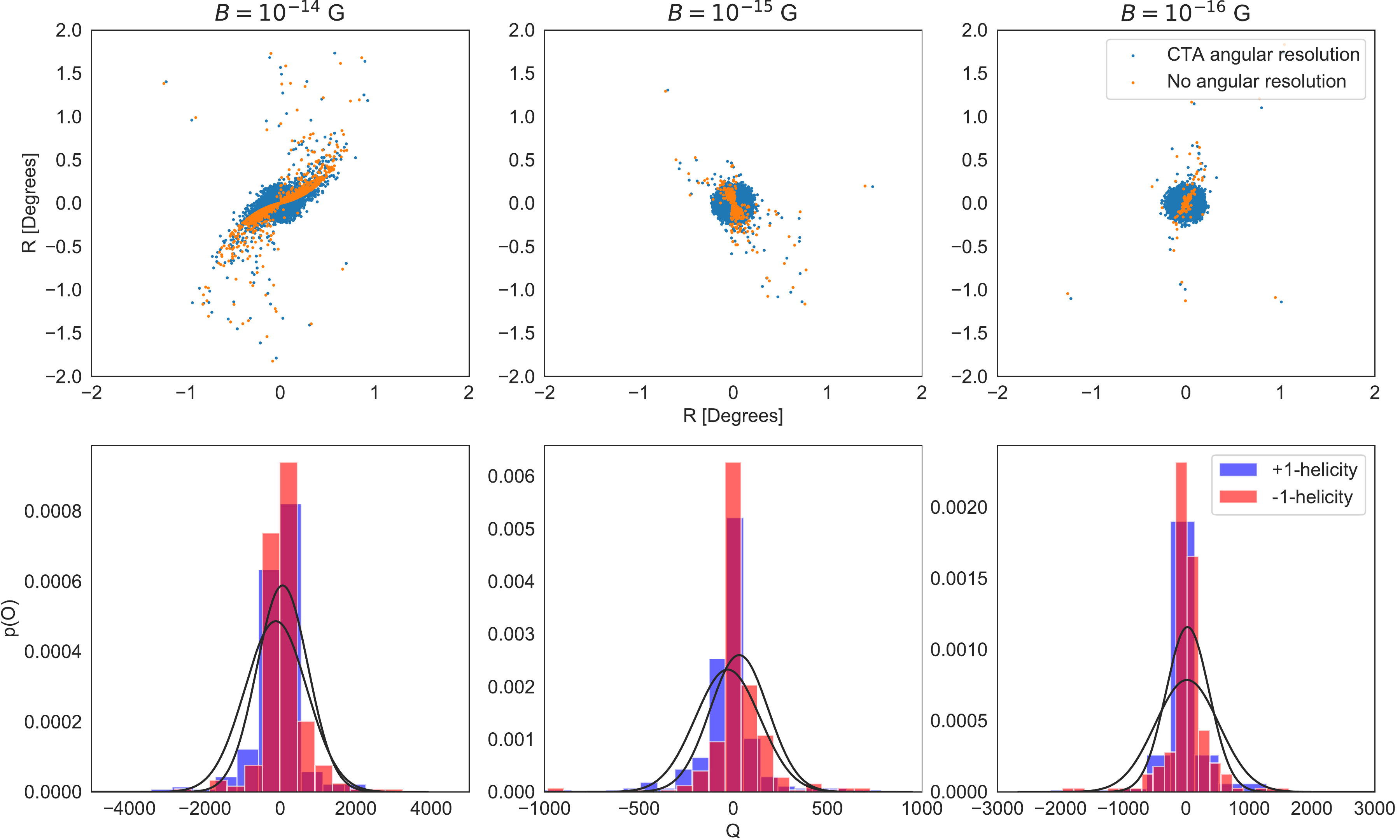}
  \caption{Top panel: Sky map of the arrival directions of photons with
    $E>100$\,GeV for three different magnetic field strengths: orange dots
    without and blue dots with accounting for the angular resolution.
    Bottom panel: MC results and fitted normal distributions of $Q$ values,
    orange bins for $h=-1$ and blue bins for $h=+1$.}
  \label{fig:CTA_1}
\end{figure*}

In order to quantify this effect, we construct the ``measured'' photon
arrival directions from the true ones by adding Gaussian noise with
variance $\sigma\simeq \ell_{68}/\sqrt{2.3}$. In Fig.~\ref{fig:CTA_1}, we show 
for three strengths of the magnetic field the effect of the finite angular
resolution of CTA on the measured arrival directions of photons with energy
$>100$\,GeV. We choose as source distance $D_s=1$\,Gpc and as wave-length
  of the magnetic field mode $L=600$\,Mpc with 10 turbulent modes.
As expected from Eq.~(\ref{theta}),
the spiral pattern is strongly blurred for $B\gsim 10^{-15}$\,G.
For $B = 10^{-14}$\,G, the blurring effect weakens, but is still too strong
for magnetic helicity to be detectable, as can be seen in the overlap plots
in the bottom of Fig.~\ref{fig:CTA_1}. Increasing the halo size further,
the signal-to-background ratio of events would decrease and more detailed
studies of specific sources taking into account their luminosity would be
needed.

%%%%%%%%%%%%%%%%%%%%%%%%%%%%%%%%%%%%%%%%%%%%%%%%%%%%%%%%%%%%%%%%%%%%%%%%%%%%%%
\section{Conclusions}                                     \label{concl}

We have searched for signatures of helical magnetic fields in more than
11\,years of gamma-ray data from Fermi-LAT. As expected from general
arguments,  we have found that this data set composed of diffuse gamma-rays is
consistent with zero helicity. We conclude that the hint for helical magnetic
fields in the 2.5\,year found in Refs.~\cite{Tashiro:2013ita,Chen:2014qva}
was a fluctuation, which statistical significance was misinterpreted
because the ``looking-elsewhere  effect'' was not accounted for.

We have also examined the detection prospects of non-zero helicity in the IGMF
using individual sources as TeV blazars. Starting from a toy model, we
have investigated how  the addition of realistic features like fluctuations
and experimental errors deteriorates the detection prospects. We have
found that charge separation can be efficiently achieved by choosing
sources with sufficiently large $\Theta_{\rm obs}$. Also the inclusion of a
distribution of magnetic field modes does not affect significantly the
signal of helical fields. In contrast, fluctuations in the
interaction lengths of the reaction $\gamma+\gamma\to e^++e^-$ and
$e^\pm+ \gamma\to e^\pm + \gamma$ together with a continuous spectrum of
injected photon energies weaken the correlation between the  energy
and the deflection angle of the observed photons and reduce thereby the
signal of helical fields considerably. However, if the halos of
tens of sources could be observed, a detection is formally still possible.

In order to quantify the detection prospects properly, more detailed
investigations taking into account both the specific experimental
properties of, e.g., CTA and the expected fluxes of TeV sources are warranted. 
Additionally, searches for more optimal estimators than the $Q$
statistics are desirable: For instance, likelihood fits of halo
templates are on general grounds known to be better, but also less robust,
estimators.

%%%%%%%%%%%%%%%%%%%%%%%%%%%%%%%%%%%%%%%%%%%%%%%%%%%%%%%%%%%%%%%%%%%%%%%%%%%%%%
\acknowledgments
\noindent
We would like to thank  Manuel Meyer, Teerthal Patel, Tanmay Vachaspati and
especially Axel Brandenburg for useful comments on this work.

%%%%%%%%%%%%%%%%%%%%%%%%%%%%%%%%%%%%%%%%%%%%%%%%%%%%%%%%%%%%%%%%%%%%%%%%%%%%%%
\subsection*{Note added}                                    

While finalizing this work, the preprint ~\cite{Asplund:2020frm} appeared.
The authors of~\cite{Asplund:2020frm} performed an analysis of the diffuse
gamma-ray data from Fermi-LAT including 11~years of data. Their analysis
is more detailed than ours and includes also a discussion of several
experimental issues of Fermi-LAT. In particular, they estimate the uncertainty
of the $Q$ statistics from Monte Carlo simulations, obtaining much larger
values than using the method of Ref.~\cite{Tashiro:2014gfa}. Adding
to this finding the ``look-elsewhere effect'' stressed in this work, the
original evidence for helical magnetic fields found in the  2.5\,year Fermi
data is weakened even more.
%They report that handedness cannot be detected in LAT data, but that “specific selection methods… could yield a significant result for $Q$.” In this paper, we study and report on particular observational constraints that would need to be made in order to detect helicity using the Tashiro method. 

%%%%%%%%%%%%%%%%%%%%%%%%%%%%%%%%%%%%%%%%%%%%%%%%%%%%%%%%%%%%%%%%%%%%%%%%%%%%%% 
%\bibliography{elmag}

\end{document}